\documentclass[aps, prb, two column,superscript address,floatfix,reprint, nobalancelastpage]{revtex4-2}
\usepackage[colorlinks=true,citecolor=blue,linkcolor=blue,urlcolor=red]{hyperref}

\newcommand{\msf}[1]{\mathsf{#1}}

\newcommand{\bsb}[1]{\boldsymbol{#1}}
\newcommand{\mbf}[1]{\mathbf{#1}}

\usepackage[sort&compress]{natbib}
\usepackage{graphicx,times}
\usepackage{color}
\usepackage{sansmath}
\usepackage{amssymb}

\usepackage{subeqnarray}
\usepackage{diagbox}
\usepackage{makecell}
\usepackage{amsmath}
\newcommand{\mathbbm}[1]{\text{\usefont{U}{bbm}{m}{n}#1}}
\newcommand{\mymathbb}[1]{\text{\usefont{U}{bbold}{m}{n}#1}}
\usepackage{tabularx}

\DeclareMathOperator{\II}{\mbf{I}}
\DeclareMathOperator{\0}{\mbf{0}}
\DeclareMathOperator{\VV}{\mbf{V}}
\DeclareMathOperator{\FF}{\msf{F}}
\DeclareMathOperator{\NN}{\msf{N}}

\begin{document}

\title{Canonical circuit quantization with linear nonreciprocal devices}

\author{A. Parra-Rodriguez}
\email{adrian.parra.rodriguez@gmail.com}
\affiliation{Department of Physical Chemistry, University of the Basque Country UPV/EHU, Apartado 644, E-48080 Bilbao, Spain}
\author{I. L. Egusquiza}
\affiliation{Department of Theoretical Physics and History of Science, University of the Basque Country UPV/EHU, Apartado 644, 48080 Bilbao, Spain}
\author{D. P. DiVincenzo}
\affiliation{Peter Gr\"unberg Institut: Theoretical Nanoelectronics, Research Center J\"ulich, D-52425 J\"ulich, Germany}
\affiliation{Institute for Quantum Information, RWTH Aachen University, D-52056 Aachen, Germany}
\author{E. Solano}
\affiliation{Department of Physical Chemistry, University of the Basque Country UPV/EHU, Apartado 644, E-48080 Bilbao, Spain}
\affiliation{IKERBASQUE, Basque Foundation for Science, Maria Diaz de Haro 3, 48013 Bilbao, Spain}
\affiliation{Department of Physics, Shanghai University, 200444 Shanghai, China}
\begin{abstract}
Nonreciprocal devices effectively mimic the breaking of  time-reversal symmetry for the subspace of dynamical variables that they couple, and can be used to create chiral information processing networks. We study the systematic inclusion of ideal gyrators and circulators into Lagrangian and Hamiltonian descriptions of lumped-element electrical networks. The proposed theory is of wide applicability in general nonreciprocal networks on the quantum regime. We apply it to pedagogical and pathological examples of circuits containing Josephson junctions and ideal nonreciprocal elements described by admittance matrices, and compare it with the more involved treatment of circuits based on nonreciprocal devices characterized by impedance or scattering matrices. Finally, we discuss the dual quantization of circuits containing phase-slip junctions and nonreciprocal devices.
\end{abstract}

\maketitle

\section{Introduction}
Low-temperature superconducting technology \cite{Devoret_2013_review} is on the verge of building quantum processors \cite{Wendin_2017_review} and simulators \cite{Barends_2015,OMalley_2016,JB_2017,Langford_2017,Kandala_2017}; machines predicted to surpass exponentially the computational power of classical computers \cite{Benioff_1982,Feynman_1982,Shor_1996,Grover_1996}. In electromagnetism, reciprocity is equivalent to the invariance of a system's linear response under interchange of sources and detectors. Nonreciprocal (NR) elements such as gyrators and circulators \cite{Tellegen_1964} have been mainly used in superconducting quantum technology as noise isolators and classical information routers, i.e., out of the quantum regime, due to the size of currently available devices. Lately, there have been several proposals for building scalable on-chip NR devices based on Josephson junction-networks \cite{Sliwa_2015_YaleRCirculator,Chapman_2017,Mueller_2018}, parametric permittivity modulation \cite{Kerckhoff_2015_BlaisCirculator}, the quantum Hall effect \cite{Viola_2014,Mahoney_2017} and mechanical resonators \cite{Barzanjeh_2017}. This nonreciprocal behavior presents quantum coherence properties \cite{Mahoney_2017} and will allow novel applications in the nontrivial routing of quantum information \cite{Lecocq_2017,Barzanjeh_2018,Metelmann_2018}. Accordingly, there is great interest in building a general framework to describe networks working fully on the quantum regime \cite{Devoret_1995_QFluct,Paladino_2003,BKD_2004,Burkard_2005,Bourassa_2012,Nigg_2013,Solgun_2014,Solgun_2015,Mortensen_2016_NM_TL,Moein_2016,Moein_2017_CutFree,Gely_2017_DivFree,ParraRodriguez_2018}. 

In this article, we use network graph theory to derive Hamiltonians of superconducting networks that contain both nonlinear Josephson junctions and ideal lineal NR devices with frequency-independent response \cite{Duinker_1959}. The correct treatment of such ideal devices will provide us with building blocks to describe more complex nonreciprocal linear devices \cite{Carlin_1964} that can be treated as linear black boxes \cite{Paladino_2003,Nigg_2013,Solgun_2014,Solgun_2015}. This theory lays the ground for the correct description of circuits in the regime where the nonreciprocal devices can be well characterized by a linear response \cite{Sliwa_2015_YaleRCirculator,Chapman_2017,Mueller_2018,Kerckhoff_2015_BlaisCirculator,Viola_2014,Mahoney_2017,Barzanjeh_2017}, even if the fundamental nonreciprocal behavior is achieved by nonlinear elements \cite{Sliwa_2015_YaleRCirculator,Chapman_2017,Mueller_2018}. Outside of this regime of validity, a black-box approach is not longer useful and a microscopic description of nonreciprocal effects is imperative. We emphasize  here that, even though they do not exist as such in nature, ideal gyrators and circulators can be useful elements to introduce in complex descriptions of networks. We focus on and extend the analyses of lumped-element networks of Devoret \cite{Devoret_1995_QFluct}, Burkard-Koch-DiVincenzo (BKD) \cite{BKD_2004}, Burkard \cite{Burkard_2005} and Solgun and DiVincenzo \cite{Solgun_2015}. Our extension involves, first, adding ideal gyrators and circulators described by an admittance ($\msf{Y}$) matrix to obtain quantum Hamiltonians with a countable number of flux degrees of freedom. As we will see, a bias towards a specific matrix description of NR devices (NRDs) appears useful when we want the Euler--Lagrange equations of motion to be current Kirchhoff equations in terms of flux variables. We next show how adding ideal NRDs described by impedance ($\msf{Z}$) or scattering ($\msf{S}$) matrices requires a more involved treatment, in that the system of equations must be first properly reduced. Finally, we also address canonical quantization with charge variables  to treat dual circuits with $\msf{Z}$-circulators; see Ref. \cite{Ulrich_2016} for a detailed description on circuit quantization with loop charges. We apply our theory to useful, pedagogical and pathological circuit examples that involve the main technical issues that more complex networks could eventually present.

Our emphasis is on quantization of an electrical network, that is to say, on quantum network \emph{analysis}, and we set aside the dual problem of network \emph{synthesis}. Even so, the introduction of the techniques presented here implies that more sophisticated synthesis methods can be used for the description of quantum devices, since our analysis can be applied to a wider class of circuits than those previously considered.

\begin{figure}[h]
	\includegraphics[width=1\linewidth]{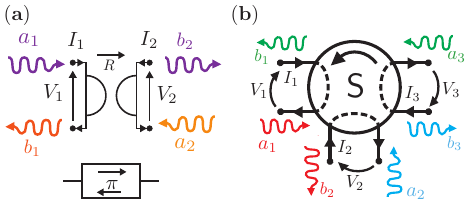}
	\caption{\label{fig:NRD_intro}({\bf a}) A 2-port gyrator: Input $a_k$ and output $b_k$ signals are related to each other through the scattering matrix $\bsb{b}=\msf{S}\bsb{a}$; with $b_2=a_{1}$ and $b_1=-a_{2}$, the element behaves as a perfect $\pi$-phase directional shifter (bottom) when impedance matched to transmission lines at ports. ({\bf b}) A 3-port circulator: Input signals transform into output signals cyclically, e.g. $b_k=a_{k-1}$. Port voltages $V_k$ and currents $I_k$ can be generally related to $a_k$ and $b_k$ through Eq. (\ref{eq:consti_SG}).}
\end{figure}

Regarding the need for a  more involved treatment of NRDs with impedance or scattering matrix presentations, bear in mind that, in microwave engineering, a multiport linear (\emph{black-box}) device can be always described by its scattering matrix parameters $\msf{S}(\omega)$ \cite{Pozar_2011}, that relate voltages and currents at its ports $\bsb{b}=\msf{S}\bsb{a}$, with $b_k=(V_k-Z_k^* I_k)/\sqrt{\mathrm{Re}\{Z_k\}}$ and $a_k=(V_k+Z_k I_k)/)/\sqrt{\mathrm{Re}\{Z_k\}}$. The reference impedances can be chosen, for simplicity, homogeneous and real, e.g., $Z_k=R\in \mathrm{Re}$. Simple properties of the scattering matrix reveal fundamental characteristics of the device. For instance, a network is reciprocal (lossless) when $\msf{S}$ is symmetric (unitary). See Fig. \ref{fig:NRD_intro} for an example of basic NR devices and their conventional symbols in electrical engineering. When ports are impedance-matched to output transmission lines ({\bf a}) a 2-port (4-terminal) ideal gyrator behaves as a perfect $\pi$-phase directional shifter, i. e. $b_2=a_{1}$ and $b_1=-a_{2}$, and ({\bf b}) a 3-port (6-terminal) ideal circulator achieves perfect signal circulation, e.g. $b_k=a_{k-1}$ \cite{Carlin_1964}. Other useful descriptions of multiport devices are the impedance $\msf{Z}(\omega)=R(1-\msf{S}(\omega))^{-1}(1+\msf{S}(\omega))$ and admittance  $\msf{Y}(\omega)=\msf{Z}^{-1}(\omega)$ matrices that relate port voltages and currents as $\VV=\msf{Z}\II$ and $\II=\msf{Y}\VV$ respectively \cite{Pozar_2011}. Although sometimes more useful, immittance descriptions of linear devices do not always exist, and working with $\msf{S}$ can be unavoidable \cite{Carlin_1964,Pozar_2011}. This comes about whenever the $\mathsf{S}$ matrix has $+1$ and $-1$ eigenvalues.

In Sec.  \ref{sec:network-graph-theory} we present some basic aspects of network graph theory, as applicable to electrical circuits, 
with reference to the current literature on its use in  quantization. We include nonreciprocal multiport elements in the consideration. We next address, in Sec. \ref{sec:networks-with-msfy}, the construction of the Lagrangian of circuits with admittance described nonreciprocal devices and the subsequent quantization. We provide specific examples of this process. In Sec.  \ref{sec:ZS_NRD} we look into the issue of nonreciprocal devices with no admittance description. To this point we have studied circuits with flux variables. In Sec. \ref{sec:dual-quant-charge} the dual, charge variables are investigated for their use in nonreciprocal circuits.  We finish with conclusions and a perspective on future work.
\section{Network Graph Theory}
\label{sec:network-graph-theory}
A lumped-element electrical network is an oriented multigraph \cite{Devoret_1995_QFluct,BKD_2004}. Each branch of the graph  connects two nodes and has a direction chosen to be that of the current passing through it. A one-port element will be  assigned a branch.   The  choice of direction for the corresponding branch is arbitrary for symmetric elements. More generally,  $N$-port elements like the \emph{circulator} are represented by $N$ branches connecting $2N$ nodes pairwise \cite{Carlin_1964}; see Fig. \ref{fig:NRD_intro}. A spanning \emph{tree} of the graph is a set of branches that connects all nodes without creating loops. The set of branches in the tree are called tree branches and all others \emph{chord} branches. Making a choice for tree and chord branches in an electrical network, we separate the currents $\mbf{I}^T=(\mbf{I}_{\mathrm{tr}}^T,\mbf{I}_{\mathrm{ch}}^T)$ and voltages $\mbf{V}^T=(\mbf{V}_{\mathrm{tr}}^T,\mbf{V}_{\mathrm{ch}}^T)$ to write Kirchhoff's equations as 
\begin{eqnarray}
\FF \II_{\mathrm{ch}}&=&-\II_{\mathrm{tr}},\label{eq:Kirchhoff_I}\\
\FF^{T}\VV_{\mathrm{tr}}&=&\VV_{\mathrm{ch}}+\dot{\mbf{\Phi}}_{\mathrm{ex}},\label{eq:Kirchhoff_V}
\end{eqnarray}
where $\FF$ is the reduced fundamental loop(/cutset) matrix describing the topology of the graph. It contains only $\{0,-1,1\}$ entries; see \cite{BKD_2004,Burkard_2005} for details on graph theory applied to superconducting circuits. Hence we make reference to $\mathsf{F}$ as the loop matrix. The vector of external fluxes $\mbf{\Phi}_{\mathrm{ex}}$ corresponds to the set of external fluxes threading each of the loops of the system.

The branch charge ($\mbf{Q}$)  and flux ($\mbf{\Phi}$) variables are defined from the flow variables $\mbf{I}$ and difference variables $\mbf{V}$ as $I_X(t)=\dot{Q}_X(t)$ and   $V_X(t)=\dot{\Phi}_X(t)$, where the subscript $X=C,\,L,\,J,\,G,\,T,\,R,\,Z,\,V,\,B$ denotes capacitors, inductors, Josephson junctions, nonreciprocal element branches, transformer branches, resistors, two-terminal impedances, voltage sources, and current sources, respectively. For the sake of simplicity we focus here on networks with passive and lossless elements, i.e., capacitors, inductors, Josephson junctions, nonreciprocal element branches, and transformer branches. We forward the reader to Refs. \cite{BKD_2004,Burkard_2005,Solgun_2014,Solgun_2015} for the inclusion of two-terminal impedances and voltage and current sources.

The constitutive equations of capacitors, inductors, and Josephson junctions are 
\begin{eqnarray}
	\mbf{Q}_C&=&\msf{C}\mbf{V}_C,\label{eq:consti_C}\\
	\mbf{I}_L&=&\msf{L}^{-1}\mbf{\Phi}_L,\label{eq:consti_L}\\
	\mbf{\Phi}_J&=&\frac{\Phi_0}{2\pi}\bsb{\varphi}_J,\label{eq:consti_J1}\\
	\mbf{I}_J&=&\mbf{I}_c\bsb{\sin}(\bsb{\varphi}_J).\label{eq:consti_J2}
\end{eqnarray}
where $\mbf{I}_c\bsb{\sin}(\bsb{\varphi}_J)$ is the column vector with $I_{ci}\sin(\varphi_{Ji})$ entries, $I_{c}$ the critical current of a junction and $\Phi_0$ the flux quantum. General multiport transformers (Belevitch transformers \cite{Belevitch_1950}) have been previously added to the Burkard analysis in Ref. \cite{Solgun_2015}. They add voltage and current constraints on the right ports in terms of its left ports and vice versa,
\begin{eqnarray}
	\II_{T}^{R}=-\NN \II_{T}^{L}, \quad \label{eq:Belevitch_trR1}
	\VV_{T}^{L}=\NN^T \VV_{T}^{R},
\end{eqnarray}
where $\msf{N}$ is the turns ratios matrix and both left and right current directions are pointing inwards. Dual transformers exist where the left-right equations (\ref{eq:Belevitch_trR1}) are inverted \cite{Belevitch_1950,Solgun_2015}. Passing now to the focus of our study, the general constitutive equation for the ideal (frequency-independent) nonreciprocal element branches can be retrieved from the scattering matrix definition
\begin{equation}
	(1-\msf{S})\VV_G(t)=(1+\msf{S})R\II_G(t),\label{eq:consti_SG}
\end{equation}
with $R$ a constant in resistance units.

In order to carry out canonical quantization in circuits, our task will be to  simplify Kirchhoff's laws together with the constitutive relations into a set of classical Euler-Lagrange (E-L) equations, from which Hamiltonian equations can be derived through a Legendre transformation, and canonically conjugate variables can be identified. In trivial cases, this reduces to having a kinetic matrix that is non-singular. 
\section{Networks with $\msf{Y}$ NRD}\label{sec:networks-with-msfy}
Given that Josephson junctions are nonlinear devices, E-L equations have been systematically derived in flux variables so as to have a  purely quadratic kinetic sector, e.g. Refs.  \cite{Devoret_1995_QFluct,BKD_2004,Burkard_2005,Solgun_2014,Solgun_2015}. In particular, BKD and Burkard quantization methods are constrained, with respect to Devoret's approach, to specific topological classes of circuits to make the Hamiltonian derivation even more systematic. For instance, in BKD all the capacitors must be included in the tree, while there are no capacitor-only loops; i.e., all capacitors are tree branches, and no external impedance can appear in the tree, while Burkard quantization has dual conditions.
These assumptions about the assignment to tree and chord branches provide us with a description of the loop matrix in block matrix form, in such a way that some of the blocks are trivial. This triviality, in turn, will allow us to construct effective loop matrices by elimination of variables.

As we shall now see, those approaches can easily incorporate ideal NR elements described by the admittance matrix ($\msf{Y}$ devices) with the realistic assumption that all of their branches are independently shunted by (parasitic) capacitors. 

For instance, the BKD formalism can be extended by assuming that all ideal NR ($G$) branches are chord branches. As stated, in BKD all capacitors of the mesh have to be in the  tree branches, whereas Josephson junctions, which are always in parallel to at least one capacitor, are chosen to be chord branches. Inductors can be both in the tree ($K$) or in the chord ($L$) set. In the following, we sketch the derivation where all inductors are chord inductors. For pedagogical purposes, we derive a Burkard circuit class extension in Appendix \ref{sec_app:Burkard_extension}. Following Ref. \cite{Solgun_2015}, we also include  Belevitch transformers in this analysis. 

The fundamental loop matrix  of a simplified BKD circuit can be written in block matrix form as
\begin{equation}
\FF=\begin{pmatrix}
\FF_{CJ}&\FF_{CL}&\FF_{CG}&\FF_{CT^{\mathrm{ch}}}\\
\FF_{T^{\mathrm{tr}}J}&\FF_{T^{\mathrm{tr}}L}&\FF_{T^{\mathrm{tr}}G}&\FF_{T^{\mathrm{tr}}T^{\mathrm{ch}}}
\end{pmatrix}.
\end{equation}
Real Josephson junctions are always in parallel to capacitors, so that $\FF_{T^{\mathrm{tr}}J}=0$. On the other hand, if all transformer left branches can be included in the tree, while transformer right branches are in the chord, then  $\FF_{T^{L}T^{R}}=\FF_{T^{\mathrm{tr}}T^{\mathrm{ch}}}=0$. We can integrate out the voltages and currents in the transformer branches \cite{Solgun_2015} inserting  (\ref{eq:Belevitch_trR1}) into Kirchhoff's equations (\ref{eq:Kirchhoff_I}, \ref{eq:Kirchhoff_V}) and write an effective loop matrix
\begin{equation}
\FF^{\mathrm{eff}}=\begin{pmatrix}
\FF_{CJ}&\FF_{CL}^{\mathrm{eff}}&\FF_{CG}^{\mathrm{eff}}
\end{pmatrix},\label{eq:BKD_Feff}
\end{equation}
with $\FF_{CL}^{\mathrm{eff}}=\FF_{CL}+\FF_{CT^{\mathrm{ch}}}\NN \FF_{T^{\mathrm{tr}}L}$ and $\FF_{CG}^{\mathrm{eff}}=\FF_{CG}+\FF_{CT^{\mathrm{ch}}}\NN \FF_{T^{\mathrm{tr}}G}$. We insert the constitutive equations (\ref{eq:consti_L}) and the admittance version of (\ref{eq:consti_SG}), $\II_G=\msf{Y}_G\VV_G$, into the reduced current equation to obtain a second-order equation in flux variables,
\begin{equation}
-\msf{C}\ddot{\bsb{\Phi}}_C=\mbf{I}_c\bsb{\sin}(\bsb{\varphi}_{C_J})+\msf{M}_0\bsb{\Phi}_C+\msf{G}\dot{\bsb{\Phi}}_C,
\end{equation}
where $\msf{M}_0=\FF_{CL}^{\mathrm{eff}}\msf{L}^{-1}(\FF_{CL}^{\mathrm{eff}})^T$, $\msf{G}=\FF_{CG}^{\mathrm{eff}}\msf{Y}_G(\FF_{CG}^{\mathrm{eff}})^T$, and $\bsb{\varphi}_{C_J}=\bsb{\varphi}_{J}$ is the vector of capacitor branch phases [related to the fluxes by (\ref{eq:consti_J1})] in parallel with the junctions. $\msf{Y}_{G}$ is a skew-symmetric matrix (because it is the Cayley transform of an orthogonal matrix $\mathsf{S}$), and by construction so is $\msf{G}$ (see Appendix \ref{sec_app:Burkard_extension}). The antisymmetry associated with the first-order derivatives, together with the fact that these second-order equations have a non-singular kinetic matrix, allows us to  derive them from the Lagrangian 
\begin{equation}	L=\frac{1}{2}\left(\dot{\bsb{\Phi}}_C^T\msf{C}\dot{\bsb{\Phi}}_C+\dot{\bsb{\Phi}}_C^T\msf{G}{\bsb{\Phi}}_C-\bsb{\Phi}_C^T\msf{M}_0{\bsb{\Phi}}_C\right)-U(\bsb{\varphi}_{C_J}).
\end{equation}
The conjugate charge variables are $\bsb{Q}_C=\partial{L}/\partial \dot{\bsb{\Phi}}_C=\msf{C}\dot{\bsb{\Phi}}_C+\frac{1}{2}\msf{G}\bsb{\Phi}_C$. Notice that conjugate charge variables are not necessarily identical to capacitor branch charge variables, which are those that appear in Eq. (\ref{eq:consti_C}). Promoting the variables to operators with canonical commutation relations $[\hat{\Phi}_{C_n},\hat{Q}_{C_m}]=i\hbar\delta_{nm}$, we derive the quantum Hamiltonian 
\begin{eqnarray}
\hat{H}=&\frac{1}{2}(\hat{\bsb{Q}}_C-\frac{1}{2}\msf{G}\hat{\bsb{\Phi}}_C)^T\msf{C}^{-1}(\hat{\bsb{Q}}_C-\frac{1}{2}\msf{G}\hat{\bsb{\Phi}}_C)\nonumber\\
&+\frac{1}{2}\hat{\bsb{\Phi}}_C^T\msf{M}_0\hat{\bsb{\Phi}}_C+U(\hat{\bsb{\varphi}}_{C_J}).\label{eq:Hamiltonian_Ydev}
\end{eqnarray}
The non linear potential is defined as $U(\hat{\bsb{\varphi}}_{C_J})=-\sum_i E_{Ji}\cos(\hat{\varphi}_{Ji})$ and the Josephson energy of each junction is $E_{Ji}=I_{ci}\Phi_0/(2\pi)$. Given the velocity-position coupling term arising from the $\msf{G}$ matrix, a form first devised in Ref. \cite{Duinker_1959}, a diagonalization of the harmonic sector requires a symplectic transformation, that can be carried out either in the classical variables or after the canonical quantization procedure; see Appendix \ref{App_sec:symplectic}. Notice the similarity of the $\mathsf{G}$ terms to a magnetic field, and their breaking time-reversal invariance. In the same manner as a magnetic field, these \emph{gyroscopic} terms are energy conserving.
\subsection*{Examples}
These extended BKD and Burkard analyses can be directly applied to a huge family of circuits to derive Hamiltonians in \emph{position}-flux variables with $\msf{Y}$ NRDs. Up till now, most of the interest in  quantization of circuits has been connected with the presence of Josephson junctions. In the present analysis we combine that presence of Josephson junctions with nonreciprocal devices. We are thus motivated to keep the flux variables as the only position coordinates of a Lagrangian/Hamiltonian mechanical system. Here we demonstrate  the quantization of two circuits consisting of two Josephson junctions coupled to (i) a general 2-port nonreciprocal black box and (ii) the specific nonreciprocal impedance response of the Viola-DiVincenzo Hall effect gyrator \cite{Viola_2014}. The first circuit is a pedagogical and useful example where the black box, in its $N$-port configuration, would represent the response of any of the given proposals in Refs. \cite{Sliwa_2015_YaleRCirculator,Chapman_2017,Mueller_2018,Kerckhoff_2015_BlaisCirculator,Viola_2014,Mahoney_2017,Barzanjeh_2017} within their valid frequency range containing two gyrators. In the second circuit, we exploit a specific 2-port impedance response, which includes a gyrator, to get an easy lumped-element approximation that can be directly quantized. Extensions of these circuits with $N$-port $\msf{Y}$ circulators would also be readily treated by this formalism. We study corner cases where the circulators cannot be described by $\msf{Y}$ matrices below in Sec. \ref{sec:ZS_NRD}.
\subsubsection{NR Black-box coupled to Josephson junctions}
\begin{figure}[h]
	\includegraphics[width=1\linewidth]{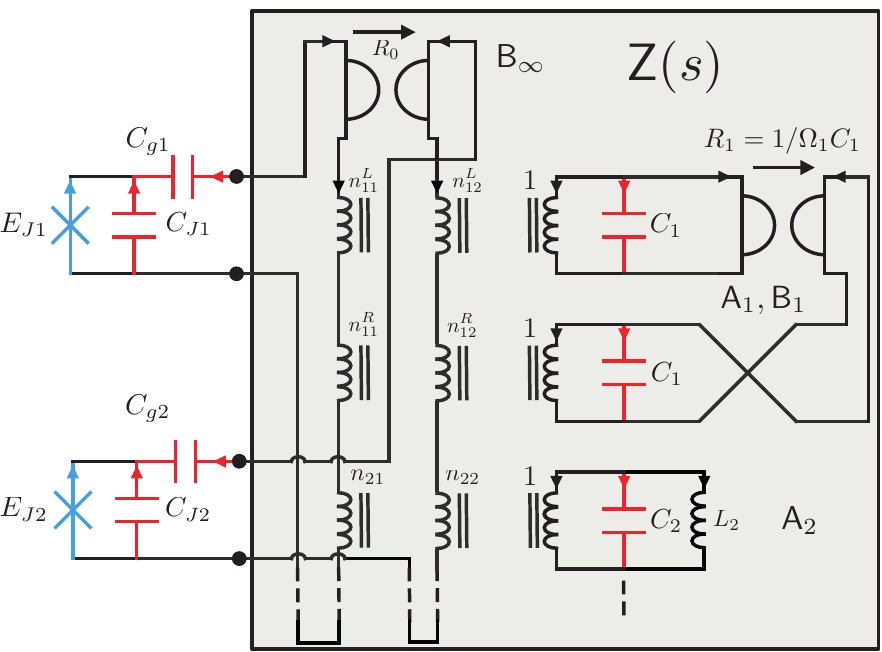}
	\caption{\label{fig:2CQ_2P_MLImpedance}Two junctions capacitively coupled to a nonreciprocal lossless impedance. Gyrator $R_0$ implements an antisymmetric pole at infinity $\msf{B}_{\infty}$. The network connected by gyrator $R_1$ yields the term $(s\msf{A}_1+\msf{B}_1)/(s^2+\Omega_1^2)$. There is a pure reciprocal stage $\msf{A}_2$. Effective tree capacitor branches are marked in red, and current directions for each branch are represented with arrows.}
\end{figure}

The first circuit consists of a 2-port nonreciprocal lossless impedance \cite{Newcomb_1966_LinearMPortSynthesis} capacitively coupled to two charge qubits at its ports; see Fig. \ref{fig:2CQ_2P_MLImpedance}. This is a generalization of the Foster reactance-function synthesis for the 1-port reciprocal  impedance $Z(s)$, with $s=i\omega$, and a simplified version of the Brune multiport impedance expansion in Refs. \cite{AndersonMoylan_1975_BruneS_SS,Solgun_2015}. 

A lossless multiport impedance matrix can be fraction-expanded as
\begin{equation}
\msf{Z}(s)=\msf{B}_{\infty}+s^{-1}\msf{A}_{0}+s\msf{A}_{\infty}+\sum_{k=1}^{\infty}\frac{s\msf{A}_{k}+\msf{B}_k}{s^{2}+\Omega_{k}^{2}}.\label{eq:NR_MportL_Zmat}
\end{equation}
It is easy to synthesize a lumped-element circuit that has this impedance to the desired level of accuracy; see  \cite{Newcomb_1966_LinearMPortSynthesis}. In a lossless linear system, the $\mathsf{S}$ matrix is unitary, and therefore $\mathsf{Z}(s)=-\mathsf{Z}^\dagger(s)$ must be anti-Hermitian. If, additionally, the system is reciprocal, it must be symmetric. The only complex parameter being $s$, a lossless reciprocal impedance matrix must be odd in the variable $s$, $-\mathsf{Z}(-s)=\mathsf{Z}(s)$. Therefore, in the fraction expansion above, the $s$-odd parts correspond to reciprocal elements, while the $s$-even parts come from non-reciprocity. Thus, all  $\msf{A}$ matrices are symmetric and are implemented by reciprocal elements while $\msf{B}$ matrices are antisymmetric, and can be decomposed into networks with gyrators. $\msf{A}_0$  and $\msf{A}_\infty$ terms  correspond to the limits $L_2\rightarrow \infty$ and $C_2\rightarrow 0$, respectively, in a reciprocal stage (see Fig. \ref{fig:2CQ_2P_MLImpedance}).   $\msf{A}_\infty$ requires special treatment, but would generally be absent because of parasitic capacitors.

The general circuit implementing $\msf{Z}(s)$ contains Belevitch transformer branches \cite{Belevitch_1950} that can be eliminated as explained above \cite{Solgun_2015} to derive a canonical Hamiltonian. An analysis of the lossless reciprocal multiport network can be found in Ref. \cite{ParraRodriguez_2018}. The tree and chord branch sets are divided in $\II_{\mathrm{tr}}^T=(\II_C^T,\II_{T^L}^T)$ and $\II_{\mathrm{ch}}^T=(\II_{J}^T,\II_L^T,\II_{T^R}^T)$, with left (right) transformer branches being tree (chord) branches. The capacitance matrix is by construction full rank and hence invertible, \begin{equation}
\msf{C}=\begin{pmatrix}
C_{J1}&&&&&&\\
&C_{J2}&&&&&\\
&&C_{g1}&&&&\\
&&&C_{g2}&&&\\
&&&&C_{1R}&&\\
&&&&&C_{1L}&\\
&&&&&&C_{2}\\
\end{pmatrix}.
\end{equation}
Inductive  $\msf{M}_0$ and gyration $\msf{G}$ matrices are computed using the turn ratios matrix 
\begin{equation}
\mathsf{N}=\begin{pmatrix}
n_{11}^L&0&0&n_{12}^L&0&0\\
0&n_{11}^R&0&0&n_{12}^R&0\\
0&0& n_{21}&0&0&n_{22}
\end{pmatrix}
\end{equation}
to calculate the effective loop submatrices $\FF_{CL}^{\mathrm{eff}}$, $\FF_{CG}^{\mathrm{eff}}$ in (\ref{eq:BKD_Feff}); see Appendix \ref{sec_app:NR_MportL} for an explicit form of the matrices. We recall that this analysis can be completed because the constitutive equation of the nonreciprocal elements (\ref{eq:consti_SG}) simplifies to $\II_G=\msf{Y}_G \VV_G$, where $\II_{G}=(I_{G0^L},I_{G0^R},I_{G1^L},I_{G1^R})^T$ and 
\begin{equation}
	\msf{Y}_G=\begin{pmatrix}
	\msf{Y}_{G0}&0\\0&\msf{Y}_{G1}
	\end{pmatrix},
\end{equation}
with $\msf{Y}_{Gi}$ the admittance matrix for each gyrator $i\in\{0,1\}$.
\subsubsection{Hall Effect NR device}
\begin{figure}[h]
	\includegraphics[width=1\linewidth]{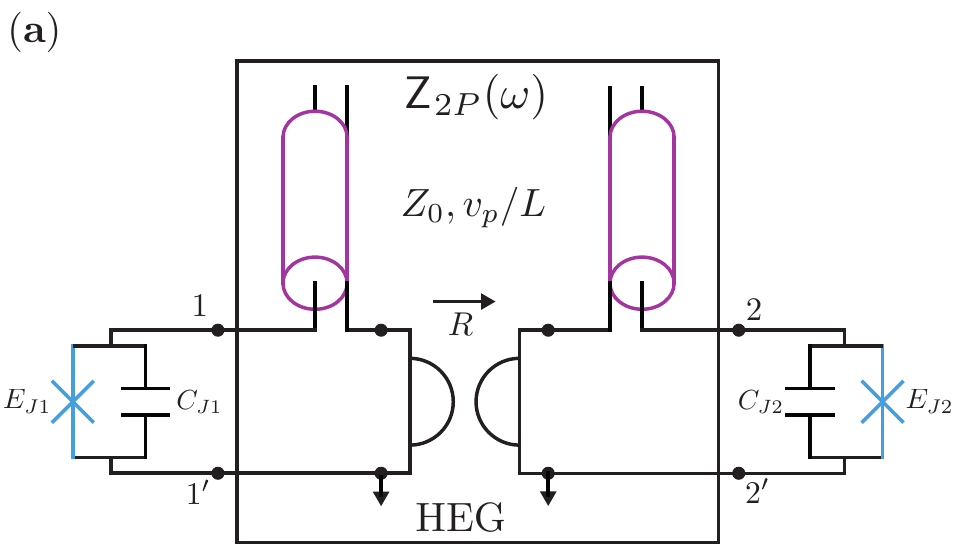}
	\includegraphics[width=1\linewidth]{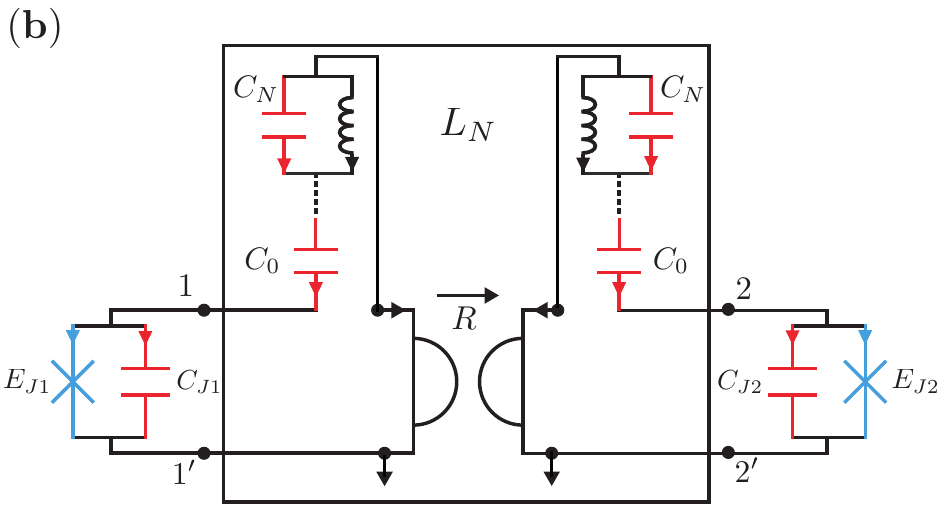}
	\caption{\label{fig:JJ_HEG_JJ} \textbf{The VD Hall effect gyrator capacitively coupled to Josephson junctions.} \textbf{({\bf a}) } An effective circuit of the device proposed by Viola and DiVincenzo matching the impedance response~(\ref{eqapp:Y_2P_DV}) consists of an ideal gyrator coupled to $\lambda/2$-transmission line stubs. ({\bf b}) The discrete approximate circuit based on a lumped element expansion of the transmission lines \cite{Pozar_2011} that is canonically quantized. Tree (capacitor) branches are marked in red.}
\end{figure}

The Hall effect has been proposed as instrumental in the implementation of  nonreciprocal devices. In Ref. \cite{Viola_2014}, capacitively coupled Hall effect devices were studied by Viola and DiVincenzo in order to break time-reversal symmetry while keeping  losses  negligible. This 2-port capacitively coupled Hall bar has an impedance matrix description \cite{Viola_2014}
\begin{equation}
\msf{Z}_{2P}(\omega)=\frac{1}{\sigma}\begin{pmatrix}
-i \cot(\omega C_L/2\sigma)&-1\\
1&-i \cot(\omega C_L/2\sigma)
\end{pmatrix},\label{eqapp:Y_2P_DV}
\end{equation}
where $\sigma$ and $C_L$ are conductance and capacitance characteristic parameters of the device, which is equivalent to that of an ideal gyrator with $R=1/\sigma$ connected in series to two $\lambda/2$-transmission line resonators of $Z_0=1/\sigma$ and $v_p/L=2\sigma/C_L$; see Fig. \ref{fig:JJ_HEG_JJ}({\bf a}). Lumped-element Foster expansions of the resonators can approximate the behavior of such a device when coupled to other lumped-element networks at its ports. This connection is achieved with lumped capacitance and inductance parameters determined by the distributed ones as $C_0=C_L/2$, $C_k=C_L/4$ and $L_k=C_L/(\sigma k \pi)^2$, for $k\in\{1,...,N\}$; see Fig. \ref{fig:JJ_HEG_JJ}({\bf b}).
We can systematically apply BKD theory and write a Lagrangian in terms of the flux branch variables of the capacitors $\bsb{\Phi}_{\mathrm{tr}}^T=\bsb{\Phi}_{C}^T=(\Phi_{C_{J_1}},\Phi_{C_{J_2}},\Phi_{0L},...,\Phi_{NL},\Phi_{0R},...,\Phi_{NR})$. The flux variables at the ports of the gyrators and at the tree capacitors are related by  $\bsb{\Phi}_{G}=\FF_{CG}^T \bsb{\Phi}_{C}$, where 
\begin{equation}
	\FF_{CG}=\begin{pmatrix}
	1&0\\
	0&1\\
	\bsb{1}_N&0\\
	0&\bsb{1}_N
	\end{pmatrix},
\end{equation}
with  $\bsb{1}_{N}$ an $N$-component column vector of ones. Explicitly, the three matrices describing the harmonic sector are the symmetric 
\begin{eqnarray}
\msf{C}&=&\begin{pmatrix}
C_{JL}&&&\\
&C_{JR}&&\\
&&\msf{C}_N&\\
&&&\msf{C}_N\\
\end{pmatrix} \quad \mathrm{and}\\
\msf{M}_0&=&\begin{pmatrix}
0&&&\\
&0&&\\
&&\msf{L}_N^{-1}&\\
&&&\msf{L}_N^{-1}\\
\end{pmatrix}
\end{eqnarray}
matrices, 
and the skew-symmetric nonreciprocal
\begin{equation}
\msf{G}=\frac{1}{R}\begin{pmatrix}
0&1&0&\bsb{1}_N^T\\
-1&0&-\bsb{1}_N^T&0\\
0&\bsb{1}_N&0&\bsb{1}_N\bsb{1}_N^T\\
-\bsb{1}_N&0&-\bsb{1}_N\bsb{1}_N^T&0
\end{pmatrix},
\end{equation}
where we have defined the capacitance submatrix $\msf{C}_N=C_0\mathrm{diag}(1,1/2,...,1/2)$ and the inductance submatrix $\msf{L}_N^{-1}=L_0^{-1}\mathrm{diag}(0,1,4,...,N^2)$,  $N$  being the number of oscillators to which we truncate the response of the resonators. Blank elements of the matrices correspond to zeros. The Hamiltonian (\ref{eq:Hamiltonian_Ydev}) can be readily computed and the canonical variables promoted to quantum operators. The diagonalization of the harmonic sector can be implemented through a symplectic transformation both before or after the quantization of variables following  Appendix \ref{App_sec:symplectic} below.
\section{Networks with $\msf{Z}$ and $\msf{S}$ NRD}
\label{sec:ZS_NRD}
The rules described above are useful to derive Hamiltonians of circuits containing ideal nonreciprocal devices characterized by a constant skew-symmetric $\msf{Y}$ matrix. However, linear systems cannot be described by admittance matrices when their $\msf{S}$ matrix has an eigenvalue $-1$. For example, ideal circulators with even (odd) number of ports,  even (odd) number of ``$-1$" entries and even (even) number of ``$1$" entries in their scattering matrix admit only $\msf{S}$-constitutive equations as in Eq. (\ref{eq:consti_SG}) (both $\msf{S}$ and $\msf{Z}$ equations) \cite{Carlin_1964}.

\begin{figure}[h]
	\includegraphics[width=1\linewidth]{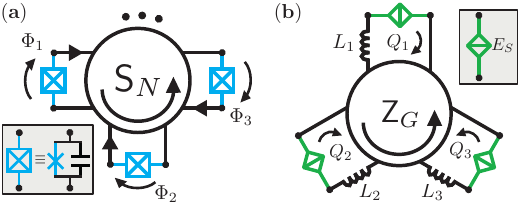}
	\caption{\label{fig:Z_S_NRCircuits} ({\bf a}) $N$-port $\msf{S}$-circulator shunted by Josephson junctions. The family of $\msf{S}$ matrices of Eq. (\ref{eq:S_N_matrix}) does not have $\msf{Y}$-description, nor does it have $\msf{Z}$ for even $N$. ({\bf b}) Dual circuit with a 3-port $\msf{Z}$-circulator shunted by phase-slip junctions in series with inductors.}
\end{figure}

We illustrate the problems arising when including circulators without $\msf{Y}$-descriptions with simple circuits containing 3- and 4-port circulators shunted by Josephson junctions; see Fig. \ref{fig:Z_S_NRCircuits}({\bf a}). Let us assume for concreteness that the $N$-port circulator is described by the scattering matrix
\begin{equation}
	\msf{S}_{N}=(-1)^N\begin{pmatrix}
	&&&1\\
	1&&&\\
	&\ddots&&\\
	&&1&
	\end{pmatrix},\label{eq:S_N_matrix}
\end{equation} 
blank elements being zero. This family of circulators cannot be assigned a $\msf{Y}$-matrix, nor do they have a $\msf{Z}$-description for even $N$. We depart from BKD and Burkard rules and choose as tree branches the circulator branches,  $\II_{\mathrm{tr}}=\II_G$, and capacitors and Josephson junction branches as chord branches $\II_{\mathrm{ch}}^T=(\II_J^T,\II_C^T)$. Kirchhoff's laws can be simply written as $-\II_G=\II_C+\II_J$ and $\VV_G=\VV_C=\VV_J=\dot{\bsb{\Phi}}$, choosing $\msf{F}_{GC}=\msf{F}_{GJ}=\mathbbm{1}$. Without loss of generality and in the interest of clarity let us assume that all Josephson junctions have homogeneous capacitances $C_{i}=C$.
Introducing Kirchhoff's and constitutive equations for capacitors ($\II_C=C \ddot{\boldsymbol{\Phi}}$) and junctions ($\II_J=\nabla_{\bsb{\Phi}}U\left(\boldsymbol{\Phi}\right)$) into (\ref{eq:consti_SG}) results in
\begin{eqnarray}
 -R\left(\mathbbm{1}+\mathsf{S}\right)(C \ddot{\boldsymbol{\Phi}}+\nabla_{\bsb{\Phi}}U(\boldsymbol{\Phi}))=\left(\mathbbm{1}-\mathsf{S}\right)\dot{\bsb{\Phi}},\label{eq:S_circuit_eom}
\end{eqnarray}
with $\nabla_{\bsb{\Phi}}U\left(\boldsymbol{\Phi}\right)=\left(U_1'(\Phi_1),U_2'(\Phi_2),...\right)^T\,$. Let $\msf{P}=\bsb{v}_{-1}\bsb{v}_{-1}^T$  be the projector onto the eigenspace of $\msf{S}$ such that $\msf{P}\msf{S}=-\msf{P}$, as it is the case for the family of matrices (\ref{eq:S_N_matrix}). Equation (\ref{eq:S_circuit_eom}) implies $\msf{P}\dot{\bsb{\Phi}}=0$; i.e., there is a frozen combination of fluxes, which corresponds to a degenerate kinetic matrix that makes the Legendre transformation impossible to perform.  A simple solution is to change coordinates to single out the frozen variable from the dynamical ones, and remove it through a projection of Eqs. (\ref{eq:S_circuit_eom}) into $\msf{Q}=\mathbbm{1}-\msf{P}$. Integrating the frozen variable, we can express $\bsb{\Phi}=\alpha\bsb{v}_{-1}+\sum_{n=1}^{N-1}\bsb{w}_n f_n$, where $\{\bsb{w}_n\}$ is a real basis expanding the projector $\msf{Q}$, $\alpha$ an initial-value flux constant, and $\{f_n\}$ the reduced set of degrees of freedom.   For the four-port case we have  the following systems of equations, $\bsb{v}_{-1}^T\dot{\bsb{\Phi}}=0$ and 
\begin{eqnarray}
C\ddot{f}_1&=&-\partial \tilde{U}_\alpha(\boldsymbol{f})/\partial f_1\label{eq:S_circuit_f2_eom}\\
C\ddot{f}_2&=&R^{-1}\dot{f}_3-\partial \tilde{U}_\alpha(\boldsymbol{f})/\partial f_2\label{eq:S_circuit_f3_eom}\\
C\ddot{f}_3&=&-R^{-1}\dot{f}_2-\partial \tilde{U}_\alpha(\boldsymbol{f})/\partial f_3\label{eq:S_circuit_f4_eom}
\end{eqnarray}
with the definition $\tilde{U}_\alpha(\boldsymbol{f})=U(\mathbf{\Phi}(\alpha,\bsb{f}))$ and $\boldsymbol{f}=\left(f_1,f_2,f_3\right)$. A similar system of equations can be derived for the three-port case except for (\ref{eq:S_circuit_f2_eom}), associated with eigenvalue $\lambda=1$ and only appearing in the four-port case; see Appendix \ref{sec:upperc-vari-circ} for the general $N$-port solution. Finally, the quantized Hamiltonian with fully dynamical variables is  
\begin{eqnarray}
\hat{H}=&\frac{1}{2C}\left({\hat{\bsb{Q}}}-\frac{1}{2}\msf{G}_{\msf{Q}}\hat{\bsb{f}}\right)^T\left(\hat{\bsb{Q}}-\frac{1}{2}\msf{G}_{\msf{Q}}\hat{\bsb{f}}\right)+\tilde{U}_\alpha(\hat{\boldsymbol{f}})\nonumber
\end{eqnarray}
with $\bsb{Q}=\partial L/\partial \dot{\bsb{f}}$ the conjugated charge variables, and the skew-symmetric matrix reads
\begin{equation}
\msf{G}_{\msf{Q}}=\frac{1}{R}\begin{pmatrix}
0&0&0\\	0&0&1\\	0&-1&0
\end{pmatrix}.\nonumber
\end{equation}
Had $-1$ not been  an eigenvalue of $\msf{S}$, all initial variables would have been dynamical. Generally, there is a coordinate transformation for any ideal circulator such that $\msf{G}$ is block diagonal, with  $2\!\!\times\!\!2$ blocks, and, possibly, one zero in the diagonal associated with  eigenvalue $+1$ (see Appendix \ref{sec:upperc-vari-circ}).

\section{Dual Quantization in Charge Variables}\label{sec:dual-quant-charge}
The procedures  explained above are useful to derive Lagrangians with flux variables as \emph{positions} in a mechanical system. Equivalent descriptions of linear systems are possible with charge-position variables, with E-L voltage equations, or with a mixed combination of both flux and charge variables. Indeed,  fluxes have been used as position variables in the context of superconducting qubits because the Josephson junction has a nonlinear current-voltage constitutive equation (\ref{eq:consti_J2}). Thus,  the Lagrangian of a circuit with these elements and $\msf{Z}$ circulators in charge variables results in nonlinear kinetic terms. Although possible, dealing with such terms is usually more cumbersome. 

In recent years, the \emph{phase-slip} (PS) junction  \cite{Mooij_2006_PSJunction,Arutyunov_2008_PSJ_2}, a nonlinear low-dissipative element in charge variable, has been implemented in superconducting technology \cite{Astafiev_2012_PSJ_exp1,Peltonen_2013_PSJ_exp2,Belkin_2015_exp4}. This element has a  constitutive equation  dual  to that of the Josephson junction; i.e., its voltage drop is  $V_{P}(t)=V_c\sin(\pi Q_{P}/e)$, and it is usually represented as in Fig. \ref{fig:Z_S_NRCircuits}({\bf b}) in green. Quantization of circuits with PS junctions and ideal $\msf{Z}$-NR elements in charge variables can be implemented directly, using the constitutive equation $\VV_G=\msf{Z}_G \II_G$. For example, the circuit in Fig. \ref{fig:Z_S_NRCircuits}({\bf b}) with a $\msf{Z}_G$-circulator, the dual circuit of Fig. \ref{fig:Z_S_NRCircuits}({\bf a}),  has the dual Lagrangian interaction term $L_{G}=(1/2)\dot{\bsb{Q}}\msf{Z}_G\bsb{Q}$ and the quantum Hamiltonian
\begin{eqnarray}
\hat{H}=&\frac{1}{2}(\hat{\bsb{\Phi}}-\frac{1}{2}\msf{Z}_G\hat{\bsb{Q}})^T\msf{L}^{-1}(\hat{\bsb{\Phi}}-\frac{1}{2}\msf{Z}_G\hat{\bsb{Q}})\nonumber+U(\hat{\bsb{Q}}),\label{eq:Hamiltonian_PS_Zdev}
\end{eqnarray}
where $\msf{L}$ is the diagonal inductance matrix and $U(\hat{\bsb{Q}})=-\sum_i E_{Si}\cos(\pi \hat{Q}_i/e)$. We forward the reader to Ref. \cite{Ulrich_2016} for a systematic quantization method of circuits with loop charges \cite{YurkeDenker_1984_QNT}.

\section{Conclusions}
We have presented a general framework to quantize canonically superconducting circuits with Josephson junctions and ideal linear nonreciprocal devices. We have introduced systematic rules for quantizing classes of circuits with ideal admittance-described nonreciprocal devices in flux variables. In such a scheme we have derived the Hamiltonian of Josephson junctions capacitively coupled to both a general linear nonreciprocal 2-port black box and the Viola-DiVincenzo gyrator at its ports. These two examples show the crucial elements that we address in the general construction, and will be of interest in their own right in forthcoming experimental devices. We have given an explicit method to quantize $N$-port ideal $\msf{Z}$ and $\msf{S}$ circulators shunted by Josephson junctions in flux variables, by careful elimination of frozen variables. Finally, we discussed an extension of these procedures to quantize circuits in terms of charge variables, a dual method of special importance when dealing with circuits containing nonreciprocal elements and phase-slip junctions. Further work will be required to add distributed elements, e.g., infinite transmission lines, to the analysis.

\begin{acknowledgments}A. P.-R. thanks Giovanni Viola, Firat Solgun and Laura Ortiz for fruitful and encouraging discussions. We acknowledge funding from Spanish MINECO/FEDER FIS2015-69983-P,  Basque Government IT986-16 and  Ph.D. Grant No. PRE-2016-1-0284. We acknowledge support from the projects QMiCS (820505) and OpenSuperQ (820363) of the EU Flagship on Quantum Technologies.
\end{acknowledgments}

\appendix
\section{\uppercase{Extended Burkard Analysis}}\label{sec_app:Burkard_extension}
We extend Burkard \cite{Burkard_2005} and Solgun-DiVincenzo \cite{Solgun_2015} analyses to include ideal multiport NR $\msf{Y}$-devices under the assumption that each branch of a NRD is shunted by a capacitor in the circuit independently. Relaxing the requirements of the BKD analysis, we allow nonreciprocal branches to appear both in the tree and in the chord set. We divide the tree and chord currents and voltages for the different components of the circuit in the following way:
\begin{eqnarray}
\mbf{\II}_{\mathrm{tr}}^T&=&\left(\II_{J}^T,\II_{L}^T,\II_{G^{\mathrm{tr}}}^T,\II_{T^{\mathrm{tr}}}^T\right),\\
\mbf{\II}_{\mathrm{ch}}^T&=&\left(\II_{C_{J}}^T,\II_{C}^T,\II_{G^{\mathrm{ch}}}^T,\II_{T^{\mathrm{ch}}}^T\right),\\
\mbf{\VV}_{\mathrm{tr}}^T&=&\left(\VV_{J}^T,\VV_{L}^T,\VV_{G^{\mathrm{tr}}}^T,\VV_{T^{\mathrm{tr}}}^T\right),\\
\mbf{\VV}_{\mathrm{ch}}^T&=&\left(\VV_{C_{J}}^T,\VV_{C}^T,\VV_{G^{\mathrm{ch}}}^T,\VV_{T^{\mathrm{ch}}}^T\right),
\end{eqnarray}
where we have added gyrator branches to  both branch sets. We can write Kirchhoff's current laws without external fluxes for simplicity, 
\begin{eqnarray}
\FF \II_{\mathrm{ch}}&=&-\II_{\mathrm{tr}},\\
\FF^{T}\VV_{\mathrm{tr}}&=&\VV_{\mathrm{ch}},
\end{eqnarray}
making use of the fundamental loop matrix $\FF$; see Refs. \cite{BKD_2004,Burkard_2005} for a detailed analysis of graph theory applied to superconducting circuits, that can be partitioned as 
\begin{equation}
\FF=\begin{pmatrix}
\mathbbm{1}&\FF_{JC}&\FF_{JG^{\mathrm{ch}}}&\FF_{JT^{\mathrm{ch}}}\\
\0&\FF_{LC}&\FF_{LG^{\mathrm{ch}}}&\FF_{LT^{\mathrm{ch}}}&\\
\0&\FF_{G^{\mathrm{tr}}C}&\FF_{G^{\mathrm{tr}}G^{\mathrm{ch}}}&\FF_{G^{\mathrm{tr}}T^{\mathrm{ch}}}&\\
\0&\FF_{T^{\mathrm{tr}}C}&\FF_{T^{\mathrm{tr}}G^{\mathrm{ch}}}&\FF_{T^{\mathrm{tr}}T^{\mathrm{ch}}}
\end{pmatrix}.
\end{equation}
We eliminate ideal transformer branches $\II_{T}^T=(\II_{T^{\mathrm{tr}}}^T,\II_{T^{\mathrm{ch}}})^T$   \cite{Solgun_2015}, which do not store energy and are not degrees of freedom of the system, by making use of Kirchhoff's current law for tree transformer branches and the current constraint equation of the transformer (\ref{eq:Belevitch_trR1}),
\begin{eqnarray}
\II_{T^{\mathrm{tr}}}&=&-(\FF_{T^{\mathrm{tr}}C}\II_{C}+\FF_{T^{\mathrm{tr}}G^{\mathrm{ch}}}\II_{G^{\mathrm{ch}}}),\\
\II_{T^{\mathrm{ch}}}&=&-\NN \II_{T^{\mathrm{tr}}},
\end{eqnarray}
with $\NN$ the turns ratios matrix. Here we have assumed that transformer tree (left) branches are not shunted by transformer chord (right) branches, i.e., $\FF_{T^{\mathrm{tr}}T^{\mathrm{ch}}}=\0$ \cite{Belevitch_1950,Solgun_2015}. We can thus express the current in the right branches of Belevitch transformer as
\begin{equation}
\II_{T^{\mathrm{ch}}}=\NN(\FF_{T^{\mathrm{tr}}C}\II_{C}+\FF_{T^{\mathrm{tr}}G^{\mathrm{ch}}}\II_{G^{\mathrm{ch}}}).
\end{equation}
We write tree Josephson, inductor, and NR tree branch currents as a function of only capacitor and NR chord branch currents,
\begin{eqnarray}
-\II_{J}&=&\II_{C_{J}}+\FF_{JC}\II_{C}+\FF_{JG^{\mathrm{ch}}}\II_{G^{\mathrm{ch}}}+\FF_{JT^{\mathrm{ch}}}\II_{T^{\mathrm{ch}}}\nonumber\\
&=&\II_{C_{J}}+\FF_{JC}^{\mathrm{eff}}\II_{C}+\FF_{JG^{\mathrm{ch}}}^{\mathrm{eff}}\II_{G^{\mathrm{ch}}}\label{eq:NR_Burkard_IJ_KKeff}\\
-\II_{L}&=&\FF_{LC}^{\mathrm{eff}}\II_{C}+\FF_{LG^{\mathrm{ch}}}^{\mathrm{eff}}\II_{G^{\mathrm{ch}}},\label{eq:NR_Burkard_IL_KKeff}\\
-\II_{G^{\mathrm{tr}}}&=&\FF_{G^{\mathrm{tr}}C}^{\mathrm{eff}}\II_{C}+\FF_{G^{\mathrm{tr}}G^{\mathrm{ch}}}^{\mathrm{eff}}\II_{G^{\mathrm{ch}}}.\label{eq:NR_Burkard_IGtr_KKeff}
\end{eqnarray}
Here, we have defined effective loop submatrices \cite{Solgun_2015}
\begin{eqnarray}
\FF_{XC}^{\mathrm{eff}}&=&\FF_{XC}+\FF_{XT^{\mathrm{ch}}}\NN\FF_{T^{\mathrm{tr}}C},\\
\FF_{XG^{\mathrm{ch}}}^{\mathrm{eff}}&=&\FF_{XG^{\mathrm{ch}}}+\FF_{XT^{\mathrm{ch}}}\NN\FF_{T^{\mathrm{tr}}G^{\mathrm{ch}}},
\end{eqnarray}
with $X=\{J,L,G^{\mathrm{tr}}\}$, that have real entries instead of the usual ternary set $\{-1,1,0\}$ for branches with currents in the same or opposite direction, or out of the loop, respectively. 

Using Kirchhoff's current law and the capacitor constitutive equation, we write the inductors in terms of the junction and inductor voltages,
\begin{eqnarray}
\II_{C_J}&=&\dot{\bsb{Q}}_{C_J}=\msf{C}_J\dot{\VV}_J,\label{eq:NR_MportL_IC1}\\
\II_C&=&\msf{C}\left((\FF_{JC}^{\mathrm{eff}})^T\dot{\mbf{V}}_J+(\FF_{LC}^{\mathrm{eff}})^T\dot{\mbf{V}}_L+(\FF_{G^{\mathrm{tr}}C}^{\mathrm{eff}})^T\dot{\mbf{V}}_{G^{\mathrm{tr}}}\right)\nonumber.\\\label{eq:NR_MportL_IC2}
\end{eqnarray}
We rewrite again current-voltage constitutive relations for inductors and junctions, Eqs. (\ref{eq:consti_L}-\ref{eq:consti_J2}) in the main text (MT), for the symmetric elements,
\begin{eqnarray}
\II_J&=&\II_c \bsb{\sin}(2\pi\bsb{\Phi}_J/\Phi_0)=-\nabla_{\bsb{\Phi}_J}U(\bsb{\Phi}_J),\label{eq:NR_Burkard_IJ_const1}\\
\II_L&=&\msf{L}^{-1}\bsb{\Phi}_L,\label{eq:NR_Burkard_IL_consti}
\end{eqnarray}
while that for the $\msf{Y}$-NR branches, Eq. (\ref{eq:consti_SG}) in the MT, can be decomposed into  
\begin{equation}
\begin{pmatrix}
\II_{G^{\mathrm{tr}}}\\	\II_{G^{\mathrm{ch}}}
\end{pmatrix}=\begin{pmatrix}
\msf{Y}_{G^{\mathrm{tr}}G^{\mathrm{tr}}}&\msf{Y}_{G^{\mathrm{tr}}G^{\mathrm{ch}}}\\	\msf{Y}_{G^{\mathrm{ch}}G^{\mathrm{tr}}}&\msf{Y}_{G^{\mathrm{ch}}G^{\mathrm{ch}}}
\end{pmatrix} \begin{pmatrix}
\VV_{G^{\mathrm{tr}}}\\	\VV_{G^{\mathrm{ch}}}
\end{pmatrix}.\label{eq:NR_Burkard_IG_consti}
\end{equation}
Introducing Kirchhoff's voltage law in the current-voltage relation for chord NR branches we derive
\begin{eqnarray}
\II_{G^{\mathrm{ch}}}&=&(\msf{Y}_{G^{\mathrm{ch}}G^{\mathrm{tr}}}+\msf{Y}_{G^{\mathrm{ch}}G^{\mathrm{ch}}}(\FF_{G^{\mathrm{tr}}G^{\mathrm{ch}}}^{\mathrm{eff}})^T)\VV_{G^{\mathrm{tr}}}\label{eq:NR_Burkard_IGch_KKeff}\\
&&+\msf{Y}_{G^{\mathrm{ch}}G^{\mathrm{ch}}}\left[(\FF_{LG^{\mathrm{ch}}}^{\mathrm{eff}})^T \VV_L+(\FF_{JG^{\mathrm{ch}}}^{\mathrm{eff}})^T \VV_J\right].\nonumber
\end{eqnarray}
Substituting Eqs. (\ref{eq:NR_Burkard_IJ_const1}-\ref{eq:NR_Burkard_IGch_KKeff}) in (\ref{eq:NR_Burkard_IJ_KKeff}, \ref{eq:NR_Burkard_IL_KKeff} and \ref{eq:NR_Burkard_IGtr_KKeff}) we have the equations of motion of the circuit that can be derived from the Lagrangian,
\begin{eqnarray}
L=\frac{1}{2}\dot{\bsb{\Phi}}^T\mathcal{C}\dot{\bsb{\Phi}}-\frac{1}{2}\bsb{\Phi}^T\msf{M}_0 \bsb{\Phi}+\frac{1}{2}\dot{\bsb{\Phi}}^T\msf{G}\bsb{\Phi}-U(\bsb{\Phi}_J)\nonumber,\\
\end{eqnarray}
with $\bsb{\Phi}^T=(\bsb{\Phi}_J^T,\bsb{\Phi}_L^T,\bsb{\Phi}_{G^{\mathrm{tr}}}^T)$. The symmetric capacitive and inductive matrices read
\begin{eqnarray}
\mathcal{C}&=&\begin{pmatrix}
\msf{C}_J & \0&\0\\\0&\0&\0\\\0&\0&\0
\end{pmatrix}+\mathcal{F}_{C}^{\mathrm{eff}}\msf{C}(\mathcal{F}_{C}^{\mathrm{eff}})^T,\\
\msf{M_0}&=&\mathcal{I}_L\msf{L}^{-1}\mathcal{I}_L^T,
\end{eqnarray}
where we defined
\begin{eqnarray}
\mathcal{F}_{X}^{\mathrm{eff}}&=&\begin{pmatrix}
\FF_{JX}^{\mathrm{eff}}\\	\FF_{LX}^{\mathrm{eff}}\\\FF_{G^{\mathrm{tr}}X}^{\mathrm{eff}}
\end{pmatrix},\quad
\mathcal{I}_L=\begin{pmatrix}
\0\\
\mathbbm{1}_{L}\\
\0\\
\end{pmatrix},
\end{eqnarray}
and $X=\{C,G^{\mathrm{ch}}\}$. The skew-symmetric matrix is
\begin{eqnarray}
\msf{G}&=&\mathcal{I}_{G^{\mathrm{tr}}}\msf{Y}_{G^{\mathrm{tr}}G^{\mathrm{tr}}}\mathcal{I}_{G^{\mathrm{tr}}}^T+\mathcal{F}_{G^{\mathrm{ch}}}^{\mathrm{eff}}\msf{Y}_{G^{\mathrm{ch}}G^{\mathrm{ch}}}(\mathcal{F}_{G^{\mathrm{ch}}}^{\mathrm{eff}})^T\label{eq:NR_Burkard_Gmat}\\
&&+\mathcal{F}_{G^{\mathrm{ch}}}^{\mathrm{eff}}\msf{Y}_{G^{\mathrm{ch}}G^{\mathrm{tr}}}\mathcal{I}_{G^{\mathrm{tr}}}+\mathcal{I}_{G^{\mathrm{tr}}}^T\msf{Y}_{G^{\mathrm{tr}}G^{\mathrm{ch}}}(\mathcal{F}_{G^{\mathrm{ch}}}^{\mathrm{eff}})^T,\nonumber
\end{eqnarray}
with the identity vector
\begin{equation}
\mathcal{I}_{G^{\mathrm{tr}}}=\begin{pmatrix}
\0\\
\0\\
\mathbbm{1}_{G^{\mathrm{tr}}}
\end{pmatrix}.
\end{equation}
Explicitly,
\begin{equation}
\msf{G}=\begin{pmatrix}
\msf{G}_{JJ}&\msf{G}_{JL}&\msf{G}_{JG^{\mathrm{tr}}}\\
-\msf{G}_{JL}^T&\msf{G}_{LL}&\msf{G}_{LG^{\mathrm{tr}}}\\
-\msf{G}_{JG^{\mathrm{tr}}}^T&-\msf{G}_{LG^{\mathrm{tr}}}^T&\msf{G}_{G^{\mathrm{tr}}G^{\mathrm{tr}}}
\end{pmatrix},
\end{equation}
where all the submatrices are defined as
\begin{eqnarray}
\msf{G}_{JJ}&=&\FF_{JG^{\mathrm{ch}}}^{\mathrm{eff}}\msf{Y}_{G^{\mathrm{ch}}G^{\mathrm{ch}}}(\FF_{JG^{\mathrm{ch}}}^{\mathrm{eff}})^T\nonumber,\\
\msf{G}_{JL}&=&\FF_{JG^{\mathrm{ch}}}^{\mathrm{eff}}\msf{Y}_{G^{\mathrm{ch}}G^{\mathrm{ch}}}(\FF_{LG^{\mathrm{ch}}}^{\mathrm{eff}})^T\nonumber,\\
\msf{G}_{LL}&=&\FF_{LG^{\mathrm{ch}}}^{\mathrm{eff}}\msf{Y}_{G^{\mathrm{ch}}G^{\mathrm{ch}}}(\FF_{LG^{\mathrm{ch}}}^{\mathrm{eff}})^T\nonumber,\\
\msf{G}_{JG^{\mathrm{tr}}}&=&\FF_{JG^{\mathrm{ch}}}^{\mathrm{eff}}(\msf{Y}_{G^{\mathrm{ch}}G^{\mathrm{tr}}}+\msf{Y}_{G^{\mathrm{ch}}G^{\mathrm{ch}}}(\FF_{G^{\mathrm{tr}}G^{\mathrm{ch}}}^{\mathrm{eff}})^T)\nonumber,\\	\msf{G}_{LG^{\mathrm{tr}}}&=&\FF_{LG^{\mathrm{ch}}}^{\mathrm{eff}}(\msf{Y}_{G^{\mathrm{ch}}G^{\mathrm{tr}}}+\msf{Y}_{G^{\mathrm{ch}}G^{\mathrm{ch}}}(\FF_{G^{\mathrm{tr}}G^{\mathrm{ch}}}^{\mathrm{eff}})^T)\nonumber,\\
\msf{G}_{G^{\mathrm{tr}}G^{\mathrm{tr}}}&=&\FF_{G^{\mathrm{tr}}G^{\mathrm{ch}}}^{\mathrm{eff}}\msf{Y}_{G^{\mathrm{ch}}G^{\mathrm{ch}}}(\FF_{G^{\mathrm{tr}}G^{\mathrm{ch}}}^{\mathrm{eff}})^T+\msf{Y}_{G^{\mathrm{tr}}G^{\mathrm{tr}}}\nonumber\\
&&+\FF_{G^{\mathrm{tr}}G^{\mathrm{ch}}}^{\mathrm{eff}}\msf{Y}_{G^{\mathrm{ch}}G^{\mathrm{tr}}}+\msf{Y}_{G^{\mathrm{tr}}G^{\mathrm{ch}}}(\FF_{G^{\mathrm{tr}}G^{\mathrm{ch}}}^{\mathrm{eff}})^T.\nonumber
\end{eqnarray}
The Hamiltonian of this system is
\begin{eqnarray}
H=&\frac{1}{2}\left(\bsb{Q}-\frac{1}{2}\msf{G}\bsb{\Phi}\right)^T\mathcal{C}^{-1}\left(\bsb{Q}-\frac{1}{2}\msf{G}\bsb{\Phi}\right)\nonumber\\
&+\frac{1}{2}\bsb{\Phi}^T\msf{M}_0 \bsb{\Phi}+U(\bsb{\Phi}_J)\label{eqapp:H_BurkardG},
\end{eqnarray}
where $\bsb{Q}=\partial L/\partial \bsb{\Phi}$ are the conjugate charges to the flux variables. Canonical quantization follows promoting the variables to operators with commutation relations $[\Phi_i,Q_j]=i\hbar$. 
\section{\uppercase{NR 2-port Impedance\\coupled to Josephson junctions}}
\label{sec_app:NR_MportL}
We explicitly compute matrices of Hamiltonian (\ref{eq:Hamiltonian_Ydev}) for circuit in Fig. \ref{fig:2CQ_2P_MLImpedance}, both in the MT, of a nonreciprocal 2-port lossless impedance \cite{Newcomb_1966_LinearMPortSynthesis} capacitively coupled to Josephson junctions. 

The tree and chord branch sets are divided in $\II_{\mathrm{tr}}^T=(\II_C^T,\II_{T^L}^T)$ and $\II_{\mathrm{ch}}^T=(\II_{J}^T,\II_L^T,\II_{T^R}^T)$, with left (right) transformer branches being tree (chord) branches. A general turns ratios matrix for the Belevitch transformer is
\begin{equation}
\mathsf{N}=\begin{pmatrix}
n_{11}^L&0&0&n_{12}^L&0&0\\
0&n_{11}^R&0&0&n_{12}^R&0\\
0&0& n_{21}&0&0&n_{22}
\end{pmatrix}.
\end{equation}
We will calculate with it the effective loop matrix (\ref{eq:BKD_Feff}) and get Hamiltonian (\ref{eq:Hamiltonian_Ydev}) in main text. The capacitance matrix is full rank
\begin{equation}
\msf{C}=\begin{pmatrix}
C_{J1}&&&&&&\\
&C_{J2}&&&&&\\
&&C_{g1}&&&&\\
&&&C_{g2}&&&\\
&&&&C_{1R}&&\\
&&&&&C_{1L}&\\
&&&&&&C_{2}\\
\end{pmatrix},
\end{equation}
where the blank elements of the matrix are zero. The inductive $\msf{M}_0$ matrix can be computed with the loop submatrix
\begin{equation}
\FF_{CL}^{\mathrm{eff}}=\FF_{CL}=
\begin{pmatrix}
\mymathbb{0}_{M}&0\\
0&1 \\
\end{pmatrix}
\end{equation}
where $M=J+g+G1+L$. $\{J,g,G1,L\}$ are, respectively, the number of (i) Josephson junctions (2), (ii) coupling capacitors (2), (iii) gyrator-shunted capacitors (2), and (iv) inductors ($L$). $\mymathbb{0}_{M}$ represents a zero square matrix of $M$ dimension. The skew-symmetric gyration matrix $\msf{G}$ can be computed using the effective loop submatrix,
\begin{eqnarray}
\FF_{CG}^{\mathrm{eff}}&=&\FF_{CG}+\FF_{CT^{R}}\NN \FF_{T^{L}G}\nonumber\\
&=&
\begin{pmatrix}
1 & 0 & 0 & 0 \\
0 & 1 & 0 & 0 \\
1 & 0 & 0 & 0 \\
0 & 1 & 0 & 0 \\
{n_{11}^L} & {n_{12}^L} & 1 & 0 \\
{n_{11}^R} & {n_{12}^R} & 0 & 1 \\
{n_{21}} & {n_{22}} & 0 & 0 \\
\end{pmatrix}
\end{eqnarray} 
which is calculated through the turn ratios matrix $\msf{N}$ and the submatrices
\begin{eqnarray}
\FF_{CG}&=&
\begin{pmatrix}
1 & 0 & 0 & 0 \\
0 & 1 & 0 & 0 \\
1 & 0 & 0 & 0 \\
0 & 1 & 0 & 0 \\
0 & 0 & 1 & 0 \\
0 & 0 & 0 & 1 \\
0 & 0 & 0 & 0 \\
\end{pmatrix},\\
\FF_{CT^{R}}&=&\begin{pmatrix}
0 & 0 & 0 \\
0 & 0 & 0 \\
0 & 0 & 0 \\
0 & 0 & 0 \\
1 & 0 & 0 \\
0 & 1 & 0 \\
0 & 0 & 1 \\
\end{pmatrix},\\
\FF_{T^{L}G}&=&
\begin{pmatrix}
1 & 0 & 0 & 0 \\
1 & 0 & 0 & 0 \\
1 & 0 & 0 & 0 \\
0 & 1 & 0 & 0 \\
0 & 1 & 0 & 0 \\
0 & 1 & 0 & 0 \\
\end{pmatrix}.
\end{eqnarray}

This analysis can be performed because the constitutive equation of the nonreciprocal elements (\ref{eq:consti_SG}) in the MT could be simplified to $\II_G=\msf{Y}_G \VV_G$, where $\II_{G}=(I_{G0^L},I_{G0^R},I_{G1^L},I_{G1^R})^T$ and 
\begin{equation}
\msf{Y}_G=\begin{pmatrix}
\msf{Y}_{G0}&0\\0&\msf{Y}_{G1}
\end{pmatrix},
\end{equation}
with
\begin{equation}
\msf{Y}_{Gi}=\frac{1}{R_i}\begin{pmatrix}
0&1\\-1&0
\end{pmatrix},
\end{equation}
the admittance matrix for each gyrator $i\in\{0,1\}$. The final gyration matrix is
\begin{equation}
\msf{G}=\FF_{CG}^{\mathrm{eff}}\msf{Y}_{G}(\FF_{CG}^{\mathrm{eff}})^T.
\end{equation}

\section{\uppercase{Symplectic diagonalization}}
\label{App_sec:symplectic}
We discuss now the procedure to diagonalize the quadratic sector of Hamiltonian (\ref{eq:Hamiltonian_Ydev}). We can perform a canonical change of variables $\bsb{Q}_C=\msf{C}^{1/2}\msf{O}^T\bsb{q}$,  $\bsb{\Phi}_C=\msf{C}^{-1/2}\msf{O}^T\bsb{f}$ such that we diagonalize the pure capacitive and inductive sectors of the Hamiltonian,
\begin{eqnarray}
H&=&\frac{1}{2}(\bsb{q}^T,\bsb{f}^T)\begin{pmatrix}
\mathbbm{1}&\msf{\Gamma}\\
\msf{\Gamma}^T&\msf{\Omega}^2
\end{pmatrix}\begin{pmatrix}
\bsb{q}\\
\bsb{f}
\end{pmatrix}+U(\bsb{f}),\label{eq:Hamiltonian_NRD_2}
\end{eqnarray}
with the definitions $\msf{\Gamma}=-\frac{1}{2}\msf{O}\msf{C}^{-1/2}\msf{G}\msf{C}^{-1/2}\msf{O}^T$ and  $\msf{\Omega}^2=\msf{O}\msf{C}^{-1/2}\msf{L}^{-1}\msf{C}^{-1/2}\msf{O}^T-\msf{\Gamma}^2$ a diagonal matrix. The conjugate variables ($\bsb{q},\bsb{f}$) are canonical in that $\{q_i,f_j\}=\delta_{ij}$. The presence of the antisymmetric matrix $\msf{\Gamma}$ in the harmonic part of the Hamiltonian leads to new normal frequencies that are greater or equal to those without it. In order to carry out canonical quantization of this Hamiltonian it is convenient to proceed with the symplectic diagonalization of the harmonic part. Consider thus the matrix
\begin{equation}
\label{eq:hamharmonicmatrix}
\msf{H}_h=\begin{pmatrix}
\mathbbm{1}&\msf{\Gamma}\\
\msf{\Gamma}^T&\msf{\Omega}^2
\end{pmatrix}\,.
\end{equation}
Since this matrix is symmetric and definite positive, the corresponding theorem of Williamson \cite{1974CeMec...9..213L} holds that it can be brought to the canonical form $\msf{D}=\mathrm{diag}\left(\msf{\Lambda},\msf{\Lambda}\right)$, with $\msf{\Lambda}$ a definite positive diagonal matrix, by a symplectic transformation $\msf{S}$. That is, $\msf{S}^T\msf{H}_h\msf{S}=\msf{D}$ with symplectic matrix $\msf{S}$. The determination of the symplectic eigenvalues and of the canonical symplectic transformation can be achieved by considering the matrix $\msf{H}_h\msf{J}$, with
\begin{equation}
\label{eq:jmatrix}
\msf{J}=\begin{pmatrix}
0&\mathbbm{1}\\
-\mathbbm{1}&0
\end{pmatrix}.
\end{equation}
Its eigenvalues form conjugate pure imaginary pairs, $\pm i \lambda_j$, where the positive numbers $\lambda_j$ are the diagonal elements of $\msf{\Lambda}$.  Choose an eigenvector $\mathbf{v}_j$ corresponding to $i\lambda_j$. Its complex conjugate, $\mathbf{v}_j^*$, is an eigenvector with $-i\lambda_j$ eigenvalue.  Organize the column eigenvectors in a matrix $\msf{F}=\begin{pmatrix}\mathbf{v}_1&\mathbf{v}_2&\cdots&\mathbf{v}_N&\mathbf{v}_1^*&\cdots&\mathbf{v}_N^*
\end{pmatrix}
$. Normalize the vectors by the condition $\msf{F}\msf{F}^\dag=\msf{H}_h$. Define a matrix function  $\msf{S}_{\msf{V}}=\left(\msf{F}^\dag\right)^{-1}\msf{V}\msf{D}^{1/2}$ acting on unitaries $\msf{V}$. It is clearly the case that, for all unitaries $\msf{V}$ and phase choices for the eigenvectors $\mathbf{v}_j$, $\msf{S}_{\msf{V}}^\dag \msf{H}_h\msf{S}_{\msf{V}}=\msf{D}$, since $\msf{F}^{-1}\msf{H}_h\left(\msf{F}^\dag\right)^{-1}=\mathbbm{1}$. The unitary $\msf{V}$ is determined by the requirement that it provide us with a symplectic matrix, $\msf{S}^T\msf{J}\msf{S}=\msf{J}$. {\it Inter alia}, this means that $\msf{S}$ is real. In fact, the choice
\begin{equation}
\label{eq:vchoice}
\msf{V}=\frac{1}{\sqrt{2}}
\begin{pmatrix}
1&i\\ 1&-i
\end{pmatrix}
\end{equation}
achieves this objective. This can be readily checked by noticing that
\begin{equation}
\label{eq:vtimesf}
\msf{V}^{\dag}\msf{F}^\dag=\frac{1}{\sqrt{2}}
\begin{pmatrix}
\mathbf{v}_1^\dag+\mathbf{v}_1^T\\
\vdots\\
\mathbf{v}_N^\dag+\mathbf{v}_N^T\\
-i\left( \mathbf{v}_1^\dag-\mathbf{v}_1^T\right)\\
\vdots   
\end{pmatrix}
\end{equation}
is explicitly real in this case, 
so $\msf{S}^{-1}=\msf{D}^{-1/2} \msf{V}^{\dag}\msf{F}^\dag$ is seen to be real. Furthermore, this choice also determines $\msf{S}$ as symplectic.

In the new variables, $\left(\boldsymbol{\xi}^T\ \boldsymbol{\pi}^T\right)=\mathsf{S}^T\left(\mathbf{q}^T\ \mathbf{f}^T\right)$, the quadratic part of the Hamiltonian is diagonal. They can now be canonically quantized, in the form $\xi_n=(a_n+a_n^\dag)/\sqrt{2}$, $\pi_n=-i(a_n-a_n^\dag)/\sqrt{2}$.    
\section{\uppercase{Reduction of variables in Circuits without $\msf{Y}$  ideal NRD}}
\label{sec:upperc-vari-circ}
We formalize and generalize the problem of the quantization of circuits in flux variables with linear NR devices that are only described by a constitutive equation through $\msf{S}$.  Further below, we apply this method to the derivation of the circuits in Fig. \ref{fig:Z_S_NRCircuits}({\bf a}) in the main text.

We start from the equation of motion (\ref{eq:S_circuit_eom}) of the main text, that we rewrite as  
\begin{eqnarray}
\label{eqapp:S_circuit_eom}
\left(\mathbbm{1}+\mathsf{S}\right)(\msf{C} \ddot{\boldsymbol{\Phi}}+\nabla_{\boldsymbol{\Phi}} U(\boldsymbol{\Phi}))=-R^{-1}\left(\mathbbm{1}-\mathsf{S}\right)\dot{\bsb{\Phi}},
\end{eqnarray}
with $\nabla_{\boldsymbol{\Phi}}U(\boldsymbol{\Phi})=\left(U_1'(\Phi_1),U_2'(\Phi_2),...\right)^T\,$, and $U_i'(\Phi_i)=E_{Ji}\sin(\Phi_i)$. $\msf{C}$ is a non-degenerate capacitance matrix. 
An ideal $N$-port circulator can always be described by a scattering matrix 
\begin{equation}
\msf{S}=\begin{pmatrix}
&&&s_N\\
s_1&&&\\
&\ddots&&\\
&&s_{N-1}&
\end{pmatrix},\label{eqapp:S_matrix}
\end{equation}
where each non-zero element can only be $s_k=\pm1$. By a correct choice of terminals, it can be proven that there are only two canonical types of ideal $N$-port circulators: those with values ($s_k=1$) in all their entries, and others with all ($s_k=1$), except for one ($s_j=-1$); see Ref. \cite{Carlin_1964} for further details.

The eigenvalue equation of the scattering matrix can be retrieved noticing that $\msf{S}^N=\prod_k s_k \mathbbm{1}$,
\begin{equation}
\lambda^N=\prod_k s_k =\pm 1.
\end{equation}
The eigenvalues of the scattering matrix lie on the unit circle, $e^{i\epsilon\pi/N} e^{2 i\pi n/N}$ with $n\in\{0,N-1\}$, and $\epsilon$ either 0 or 1. The eigenvalue $\lambda=-1$ appears with multiplicity one for $N$ even ($N$ odd)  with $\prod s_k=1$ ($\prod s_k=-1$). On the other hand, the eigenvalue $\lambda=1$ is present also with multiplicity one for $N$ both even and odd  when $\prod s_k=+1$.  All other eigenvalues come in pairs of complex conjugate values ($\lambda_k$ and $\lambda_k^*$).

Let us assume that $\msf{S}$ presents eigenvalue $-1$. We define the projector $\msf{P}=\bsb{v}_{-1}\bsb{v}_{-1}^T$ such that $\msf{S}\msf{P}=-\msf{P}=\msf{P}\msf{S}$, where $\bsb{v}_{-1}$ is the normalized eigenvector corresponding to the eigenvalue $-1$. We complete the identity with the projector $\msf{Q}=\mathbbm{1}-\msf{P}$, which also commutes with $\msf{S}$; $[\msf{S},\msf{Q}]=[\msf{S},\msf{P}]=0$. It is trivial to prove that $\msf{P}$ is real and that thus so it is $\msf{Q}$. If $-1$ is an eigenvalue, it always has multiplicity 1. Then, given that $\msf{S}=\msf{S}^*$,
\begin{eqnarray}
\left(\msf{S}\bsb{v}_{-1}\right)^*&=&-\bsb{v}_{-1}^*=\msf{S}\bsb{v}_{-1}^*,\\
\msf{S}\bsb{v}_{-1}&=&-\bsb{v}_{-1}.
\end{eqnarray}
The above two equations can only be true if $\bsb{v}_{-1}=\bsb{v}_{-1}^*$. 
Then, applying $\msf{P}$ to Eq. (\ref{eqapp:S_circuit_eom}), we have
\begin{equation}
\msf{P}\dot{\bsb{\Phi}}=0.
\end{equation}
This equation can be integrated, so that the flux variable vector is expressed as
\begin{equation}
\bsb{\Phi}=\msf{P}\bsb{\Phi}+\msf{Q}\bsb{\Phi}=\alpha \bsb{v}_{-1}+ \bsb{\Psi},\label{appeq:phi_exp}
\end{equation}
where we defined $\bsb{\Psi}=\msf{Q}\bsb{\Phi}$, and $\alpha$ is an initial-value constant in flux units. Inserting the above expression in the equation of motion and applying $\msf{Q}$ on the left, we have
\begin{eqnarray}
\msf{Q}\left(\mathbbm{1}+\mathsf{S}\right)\msf{Q}(\msf{C}_{\mathsf{Q}} \ddot{\boldsymbol{\Psi}}+\msf{Q}\nabla_{\boldsymbol{\Psi}}\tilde{U}_\alpha(\boldsymbol{\Psi}))=\nonumber
-R^{-1}\msf{Q}\left(\mathbbm{1}-\mathsf{S}\right)\msf{Q}\dot{\bsb{\Psi}},\nonumber\\
\end{eqnarray}
with $\msf{C}_{\msf{Q}}=\msf{Q}\msf{C}\msf{Q}$  a new symmetric reduced capacitance matrix, and $\tilde{U}_\alpha(\boldsymbol{\Psi})=U(\msf{Q}\boldsymbol{\Phi}+\alpha\bsb{v}_{-1})$ the new potential. The differential nabla operator on the original flux variables becomes  $\nabla_{\bsb{\Phi}}=\msf{Q}\nabla_{\bsb{\Psi}}+\bsb{v}_{-1}\partial_{\alpha}$. In this new $N-1$ dimensional space, the remnant of $\mathsf{Q}\left(\mathbbm{1}+\mathsf{S}\right)\mathsf{Q}$  is invertible. Formally, we derive in this reduced space the Euler-Lagrange equation 
\begin{eqnarray}
\mathsf{C}_{\msf{Q}} \ddot{\boldsymbol{\Psi}}+\msf{Q}\nabla_{\boldsymbol{\Psi}}\tilde{U}_\alpha(\boldsymbol{\Psi})=
-\msf{G}_{\msf{Q}}\dot{\bsb{\Psi}},\label{appeq:S_circuit_EL_formal}
\end{eqnarray}
with $\msf{G}_{\msf{Q}}=R^{-1}(\msf{Q}\left(\mathbbm{1}+\mathsf{S}\right)\msf{Q})^{-1}(\msf{Q}\left(\mathbbm{1}-\mathsf{S}\right)\msf{Q})$, again understood in the reduced space. There,  $\mathsf{G}_{\mathsf{Q}}$ is the Cayley transform of an orthogonal matrix, and thus a skew-symmetric matrix. 

Let us illustrate the procedure with the choice of a specific decomposition of the real projector $\msf{Q}$. Consider   $\bsb{v}_k$ and its complex conjugate $\bsb{v}_k^*$ to be orthogonal vectors in the subspace complementary to $\msf{P}$. It is  then easy to prove that real $\mathrm{Re}\{\bsb{v}_k\}=(\bsb{v}_k+\bsb{v}_k^*)/2$ and imaginary parts $\mathrm{Im}\{\bsb{v}_k\}=-i(\bsb{v}_k-\bsb{v}_k^*)/2$ are orthogonal vectors, again orthogonal to the $\msf{P}$ eigenspace. This assumption will hold if the vector $\bsb{v}_k$ is an eigenvector of $\mathsf{S}$ with complex eigenvalue. If the eigenvalue $\lambda=1$ is present, its associated eigenvector is also real; the proof is completely analogous to the above for the eigenvector $\bsb{v}_{-1}$. Normalizing all vectors, we can write
\begin{eqnarray}
\msf{Q}&=\bsb{v}_{1}\bsb{v}_{1}^T+\sum_k \bsb{x}_k \bsb{x}_k^T + \bsb{y}_k \bsb{y}_k^T,\nonumber\\
&=\sum_{n=1}^{N-1} \bsb{w}_n \bsb{w}_n^T,\label{appeq:Z_S_Q_w}
\end{eqnarray}
with $\bsb{x}_k=\mathrm{Re}\{\bsb{v}_k\}/||\mathrm{Re}\{\bsb{v}_k\}||$ and $\bsb{y}_k=\mathrm{Im}\{\bsb{v}_k\}/||\mathrm{Im}\{\bsb{v}_k\}||$,  $k$ running through all the vectors coming in complex conjugate pairs. In general, let us denote by  $\bsb{w}_n$ those real orthonormal vectors spanning the orthogonal space.

Using this nomenclature and Eq. (\ref{appeq:phi_exp}) we write 
\begin{eqnarray}
\bsb{\Phi}&=\alpha \bsb{v}_{-1}+\sum_n f_n \bsb{w}_n,\nonumber\\
&=\msf{M}\begin{pmatrix}
\alpha\\
\bsb{f}
\end{pmatrix},\label{appeq:Z_S_phi_M_f}
\end{eqnarray}
with $f_n=\bsb{w}_n^T\bsb{\Phi}$, and $\msf{M}=[\bsb{v}_{-1},\bsb{w}_1,\bsb{w}_2,...]$ an orthogonal matrix, i.e., $\msf{M}\msf{M}^T=\mathbbm{1}$. The nabla operator can be rewritten as 
\begin{equation}
\nabla_{\bsb{\Phi}}=(\msf{M}^{-1})^T\begin{pmatrix}
\frac{\partial}{\partial \alpha}\\
\frac{\partial}{\partial f_1}\\
\vdots
\end{pmatrix}=\msf{M}\begin{pmatrix}
\frac{\partial}{\partial \alpha}\\
\nabla_{\bsb{f}}
\end{pmatrix}.\label{appeq:Z_S_nabla_phi_f}
\end{equation}
Finally, inserting the above decompositions  (\ref{appeq:Z_S_phi_M_f},\ref{appeq:Z_S_nabla_phi_f},\ref{appeq:Z_S_Q_w}) in Eq. (\ref{appeq:S_circuit_EL_formal}), we  rewrite the equation of motion
\begin{eqnarray}
\sum_{n,m,l} \bsb{w}_n (\mathbbm{1}+\msf{S})_{nm}\left[ (\msf{C})_{ml}\ddot{f}_l+\partial_{f_m}\tilde{U}_{\alpha}(\bsb{f})\right]\nonumber\\
=-R^{-1}\sum_{n,l}  \bsb{w}_n(\mathbbm{1}-\msf{S})_{nl} \dot{f}_l,
\end{eqnarray}
with $(\msf{A})_{rt}=\bsb{w}_r^T\msf{A}\bsb{w}_t$, together with $\dot{\alpha}=0$. Multiplying from the left with the real row vectors $\{\bsb{w}_n^T\}$,  and inverting the first matrix on the left-hand side, we arrive at an explicit form of Eq. (\ref{appeq:S_circuit_EL_formal})
\begin{equation}
(\msf{C})_{ml}\ddot{f}_l+\partial_{f_m}\tilde{U}_{\alpha}(\bsb{f})
=-(\msf{G}_{\msf{Q}})_{ml}\dot{f}_l,\nonumber\\
\end{equation}
where we have defined $(\msf{G}_{\msf{Q}})_{ml}=R^{-1}(\mathbbm{1}+\msf{S})_{mn}^{-1}(\mathbbm{1}-\msf{S})_{nl}$ and we have used Einstein's notation of summation over repeated indices. Here, we can identify $\bsb{\Psi}\equiv(0,\bsb{f})$ in Eq.~(\ref{appeq:S_circuit_EL_formal}). Furthermore, the matrix $\mathsf{C}_{\mathsf{Q}}$ has as matrix elements in this basis precisely $ (\msf{C})_{ml}$. The Lagrangian without constraints and full-rank kinetic matrix with such equations of motion  
\begin{eqnarray}
L=&\frac{1}{2}\dot{\boldsymbol{f}}^T\msf{C}_{\msf{Q}}\dot{\boldsymbol{f}}+\frac{1}{2}\dot{\boldsymbol{f}}^T \msf{G}_{\msf{Q}}\boldsymbol{f}-\tilde{U}_{\alpha}(\boldsymbol{f}),\nonumber
\end{eqnarray}
with $\boldsymbol{f}=\left(f_1,f_2,...\right)$. Finally, the quantized Hamiltonian is  
\begin{eqnarray}
\hat{H}=&\frac{1}{2}\left({\hat{\bsb{Q}}}-\frac{1}{2}\msf{G}_{\msf{Q}}\hat{\bsb{f}}\right)^T\msf{C}_{\msf{Q}}^{-1}\left(\hat{\bsb{Q}}-\frac{1}{2}\msf{G}_{\msf{Q}}\hat{\bsb{f}}\right)+\tilde{U}_\alpha(\hat{\boldsymbol{f}}),\nonumber
\end{eqnarray}
again with  $\bsb{Q}=\partial L/\partial \dot{\bsb{f}}$ the conjugated charge variables, later to be promoted to operators.
\subsection*{Examples}
Let us now use this general theory to quantize the specific cases illustrated in the main text. The scattering matrix of Eq. (15) introduced in the circuits in Fig. 3({\bf a}) in the MT,
\begin{equation}
\msf{S}_{N}=(-1)^N\begin{pmatrix}
&&&1\\
1&&&\\
&\ddots&&\\
&&1&
\end{pmatrix},\label{eqapp:S_N_matrix}
\end{equation} 
has $-1$ eigenvalues for all $N$ and $+1$ eigenvalues for even-$N$ numbers of ports. Notice that in the analysis of the equations of motion above we have not made use of the canonical form of $\mathsf{S}$ matrices mentioned earlier, and indeed this example does not and needs not conform to that canonical presentation.

\subsubsection{3-port case}
The eigenvalues and eigenvectors for $N=3$ are  $\bsb{\lambda}_3=(-1,\lambda_3,\lambda_3^*)$ and $\msf{V}_3=[\bsb{v}_{-1},\bsb{v}_3,\bsb{v}_3^*]^T$, respectively, with $\lambda_3=e^{2\pi i/3}$ and $\bsb{v}_3=(e^{2\pi i/3},e^{-2\pi i/3},1)/\sqrt{3}$.  The eigenvalue $\lambda=-1$ of $\msf{S}_N$, present in this family of matrices, is associated with the constraint $\bsb{v}_{-1}^T\dot{\bsb{\Phi}}=0$ where $\bsb{v}_{-1}=(1,1,1)/\sqrt{3}$ is the normalized eigenvector.

We can apply the theory described above to compute the projectors 
\begin{eqnarray}
\msf{P}&=&\bsb{v}_{-1}\bsb{v}_{-1}^T=\frac{1}{3}\begin{pmatrix}
1&1&1\\
1&1&1\\
1&1&1
\end{pmatrix},\\
\msf{Q}&=&\mathbbm{1}-\msf{P}=\frac{1}{3}\begin{pmatrix}
2 & -1 & -1 \\
-1 &2 & -1 \\
-1 & -1 & 2
\end{pmatrix}.
\end{eqnarray}
The reduced capacitance matrix is 
\begin{equation}
\msf{C}_{\msf{Q}}=\frac{1}{2}\left(
\begin{array}{cc}
\frac{1}{3} \left(C_1+C_2+4 C_3\right) & \frac{C_2-C_1}{ \sqrt{3}} \\
\frac{C_2-C_1}{ \sqrt{3}} &  \left(C_1+C_2\right) \\
\end{array}
\right),
\end{equation}
while the gyration matrix is
\begin{equation}
\msf{G}_{\msf{Q}}=\frac{1}{R\sqrt{3}}\begin{pmatrix}
0&1\\
-1&0
\end{pmatrix}.
\end{equation}
Finally, the potential function $\tilde{U}_\alpha=U(\msf{M}(\alpha,\bsb{f}^T)^T)$, with $U(\bsb{\Phi})=-\sum_{i=1}^{3}E_{Ji}\cos(\Phi_i)$ and 
\begin{equation}
\msf{M}=\begin{pmatrix}
\frac{1}{\sqrt{3}} & \frac{1}{\sqrt{3}} & \frac{1}{\sqrt{3}} \\
-\frac{1}{\sqrt{6}} & -\frac{1}{\sqrt{6}} & \sqrt{\frac{2}{3}} \\
\frac{1}{\sqrt{2}} & -\frac{1}{\sqrt{2}} & 0 
\end{pmatrix}.
\end{equation}
\subsubsection{4-port case}
The eigenvalues and eigenvectors for $N=4$ are  $\bsb{\lambda}_4=(-1,1,\lambda_4,\lambda_4^*)$  and $\msf{V}_4=[\bsb{v}_{-1},\bsb{v}_{1},\bsb{v}_4,\bsb{v}_4^*]^T$, respectively, with $\lambda_4=i$ and $\bsb{v}_{4}=(-i,-1,i,1)/2$.  The eigenvalue $\lambda=-1$ of $\msf{S}_N$, present in this family of matrices, is associated with the constraint $\bsb{v}_{-1}^T\dot{\bsb{\Phi}}=0$ where $\bsb{v}_{-1}=(-1,1,-1,1)/2$ is the normalized eigenvector.

The inhomogeneous capacitance matrix is 
\begin{equation}
\msf{C}_{\msf{Q}}=\left(
\begin{array}{ccc}
\frac{C_1+C_2+C_3+C_4}{4} & \frac{C_4-C_2}{2 \sqrt{2}} & \frac{C_3-C_1}{2 \sqrt{2}} \\
\frac{C_4-C_2}{2 \sqrt{2}} & \frac{C_2+C_4}{2}  & 0 \\
\frac{C_3-C_1}{2 \sqrt{2}} & 0 & \frac{C_1+C_3}{2}  \\
\end{array}
\right),
\end{equation}
that reduces to $\msf{C}_{\msf{Q}}=C\mathbbm{1}$ for $C_i=C$. On the other hand, the gyration matrix has now a zero column and row corresponding to the eigenvalue $\lambda=+1$,
\begin{equation}
\msf{G}_{\msf{Q}}=\frac{1}{R}\begin{pmatrix}
0&0&0\\	0&0&1\\	0&-1&0
\end{pmatrix}.
\end{equation}
Given the complex-conjugate pairwise nature of the eigenvalues and eigenvectors, the gyration matrix can always be written in a basis with $2\!\!\times\!\!2$ blocks, except for the  row and the column of zeros corresponding to the $+1$ eigenvalue. Finally, we have the potential function $\tilde{U}_\alpha=U(\msf{M}(\alpha,\bsb{f}^T)^T)$, with $U(\bsb{\Phi})=-\sum_{i=1}^{3}E_{Ji}\cos(\Phi_i)$ and 
\begin{equation}
\msf{M}=\begin{pmatrix}
-\frac{1}{2} & \frac{1}{2} & -\frac{1}{2} & \frac{1}{2} \\
\frac{1}{2} & \frac{1}{2} & \frac{1}{2} & \frac{1}{2} \\
0 & -\frac{1}{\sqrt{2}} & 0 & \frac{1}{\sqrt{2}} \\
-\frac{1}{\sqrt{2}} & 0 & \frac{1}{\sqrt{2}} & 0
\end{pmatrix}.
\end{equation}
\\
 
\end{document}